# A Novel Single-Mode Microwave Assisted Synthesis of Metal Oxide as Visible-light Photocatalyst

Kunihiko Kato[a], Sebastien Vaucher [a,b], Patrik Hoffmann [a,b], Yunzi Xin [a], Takashi Shirai [a*]

[a] *Advanced Ceramics Research Center, Nagoya Institute of Technology, Gokiso, Showa-ku, Nagoya, Aichi 466-8555 Japan*
[b] *Swiss federal laboratories for materials science and technology (Empa), Feuerwerkerstrasse 39, Thun CH-3602 Switzerland*

*Corresponding authors: e-mail: shirai@nitech.ac.jp*

## Abstract

Visible-light photocatalyst titanium dioxide ($TiO_2$) was successfully prepared via a novel and facile single-mode microwave assisted synthesis process. In this one-step synthesis, Ti as target material selectively oxides in magnetic field throughout rapid heating, whose process requires less energy consumption and short time. In obtained $TiO_2$, self-doping of $Ti^{3+}$ was confirmed, which makes $TiO_2$ performed sufficient light absorption in visible region with wavelength above 400 nm. Such $Ti^{3+}$ self-doped $TiO_2$ exhibits much narrower optical bandgap (2.14 eV) to compare with stoichiometric $TiO_2$ (3.0-3.2 eV). The synthesized $TiO_2$ also shows superior photocatalytic activity to commercially available $TiO_2$ towards the degradation of Rhodamine B under visible light irradiation.

## 1. Introduction

Titanium dioxide ($TiO_2$) is one of the most widely utilized photocatalysis with respects to its physical and chemical stability, high photocatalytic activity, and nontoxicity [1,2]. However, the wide bandgap (3.0 - 3.2 eV) of $TiO_2$ seriously limits its photocatalytic performance only in UV light region, which is about only 5 % of incoming solar spectrum. The development of $TiO_2$ with activated photocatalysis under visible light irradiation is important for practical application. In order to improve the limited optical absorption of $TiO_2$, many efforts have been made to optimize the band structure of $TiO_2$ by inducing donor or acceptor states via doping of metals [3,4] or nonmetals impurities [5-8]. It has been also reported that inducing of oxygen vacancies and self-doping of $Ti^{3+}$ into stoichiometric $TiO_2$ effectively enhanced the light absorption in visible region and photocatalytic activities. Several approaches have been reported for the preparation of oxygen-deficient and $Ti^{3+}$ doped $TiO_2$, such as sol-gel processing, hydrothermal/solvothermal synthesis and magnetron sputtering method and so on. However, these processes require harsh experimental conditions, such as inert atmosphere, high temperature heating and high-pressure operation [9-11].

Microwave (MW) has been widely utilized as rapid heating process for past several decades, according to the direct interaction between target material and electro (E)- / magnetic (H)- field involved in MW and higher efficiency compared to conventional heating process. The absorbing properties of material in electromagnetic filed depend on its physical and chemical properties. Namely, metals perform good absorbing properties in H-filed while oxides perform absorbing properties in E-field[12-15]. Therefore, single-mode microwave applicator which can separate E- and H- filed components effectively has attracted much attention for selectively heating of different target materials [12-15].

In this work, we report a novel and facile single-mode MW assisted synthesis process of $Ti^{3+}$ doped $TiO_2$. Upon the one-step irradiation of single-mode MW on titanium (Ti) target in oxygen, $Ti^{3+}$ doped $TiO_2$ was successfully prepared in tens of second reaction via rapid temperature change and short heat history, which can be attributed to the drastic change of MW absorbing properties accompanied with change of chemical state in target material. To the best of our knowledge, the present work should be the first report on preparation of functional metal oxide from metal target via single-mode MW assisted synthesis. In addition, the obtained $Ti^{3+}$ doped $TiO_2$ performs sufficient light absorption in visible region and exhibits narrow band gap of 2.14 eV as well as superior photocatalytic activity in degradation of Rhodamine B under visible light irradiation. The innovative synthesis process of present work provides a brand-new direction of synthesis routine for metal oxides with specific chemical and physical properties.

## 2. Materials and methods

As for the synthesis of $TiO_2$, $TE_{111}$ mode 2.45 GHz cylindrical MW cavity was used in MW heating. 1.0 g of Ti powder (3N, powder under 45 μm mesh, Kojundo Chemical Laboratory, Japan), as target material, was loaded into a quartz tube and heated up in oxygen atmosphere under H-Field at fixed output of 20 W. During MW heating, the resonance frequency and temperature of the sample were measured by Vector Network Analyzer (VNA) and radiation pyrometer, respectively. The temperature and resonance freaquency during MW heating are shown in Fig. S1.

The crystal structure of synthesized $TiO_2$ (hereinafter named as "MWS"), raw Ti powder and two kinds of commercial $TiO_2$ powders (rutile (Kojundo, Japan) and P-25 (Degussa), respectively) as references were analyzed by X-ray diffraction (XRD: Ultima IV, Rigaku,

Japan) with Cu-Kα line. The chemical structure of synthesized sample was also investigated by Raman spectra (NRS-3100, Jasco, Japan). The morphology and crystallinity of $TiO_{2-x}$ on Ti particle were investigated by transmission electron microscopy (TEM: JEM-ARM200F, JEOL, Japan) and electron diffraction pattern. Surface chemical state of Ti in sample was analyzed by X-ray photoelectron spectroscopy with Al Kα X-rays radiation (hv = 1486.6 eV) (XPS: M-prove, SSI, USA). UV and visible light absorption spectra were characterized by a commercial UV-Vis spectrophotometer (V-7100, Jasco, Japan).

As for photocatalytic degradation of Rhodamin B, a 500 W xenon lamp was used as the visible light source. Rhodamine B solution (20 ml) and the catalysis (MWS and S2) were taken in a beaker and exposed to visible light for up to 180 min. The solution samples of about 2-3 ml were taken out at a regular interval from the test solution, centrifuged for 5 min at 10000 rpm and their absorbance were recorded at 555 nm using a spectrometer (V-7100, Jasco, Japan). The decomposition amount of Rhodamine B was calculated as following formula.

$$\Delta C = c_i - c_0 \qquad (1)$$

Here, $\Delta C$ is the decomposition amount of Rhodamine B (ppm), $c_i$ and $c_0$ are the concentration of Rhodamine B in the solution after being exposed and being reached equilibrium in dark. The concentration was estimated from the calibration curve using absorbance of Rhodamine B by Beer-Lambert law.

## 3. Results and Discussion

The XRD patterns of synthesized $TiO_2$, raw Ti, rutile and P-25 $TiO_2$ are shown in Fig.1. It can be confirmed that rutile $TiO_2$ phase (JCPDS-ICDD card No. 01-084-1283) was observed in synthesized sample while unreacted Ti was also remained. As Raman spectra shown in Fig. S2, the characteristic Raman-active modes of the rutile $TiO_2$ phase was also observed with bands appear at 143 cm$^{-1}$ ($B_{1g}$), 250 cm$^{-1}$ (multi-phonon process), 420 cm$^{-1}$ ($E_g$) and 612 cm$^{-1}$ ($A_{1g}$) [16]. No signals corresponding to any other crystallized $TiO_2$ was detected.

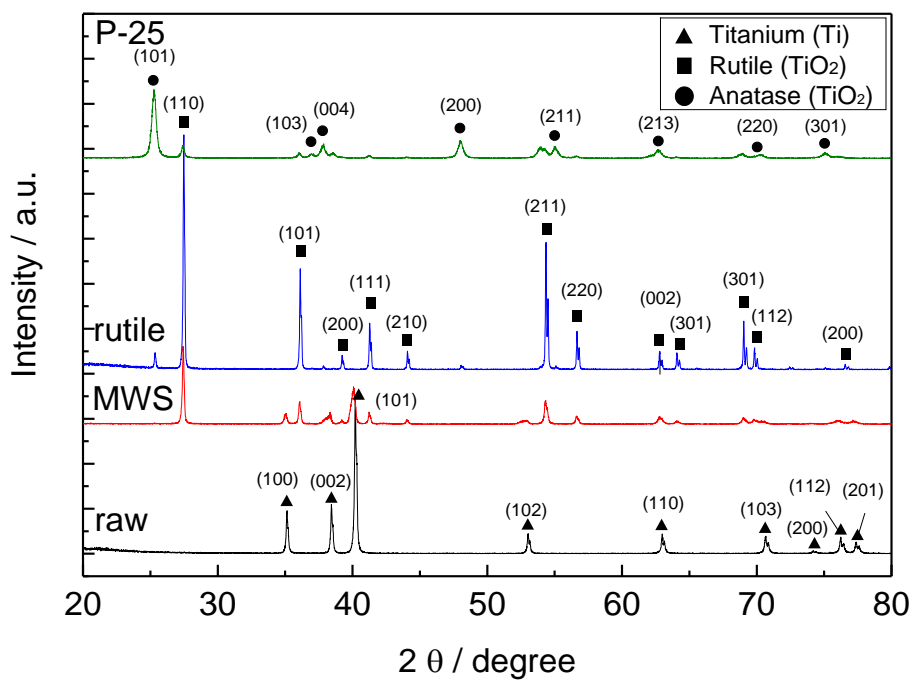

**Fig. 1.** The XRD pattern of raw and obtained material.

Fig.2 displays the TEM images, selected area electron diffraction pattern and EDS mapping results of synthesized $TiO_2$. The EDS mapping (Fig.2 (c-e)) results of the elemental distribution of Ti and O indicates that part of O was only distributed in surface area. It illustrates that the synthesized sample exhibits core/shell structure of Ti/ $TiO_2$ with 2.8 μm thickness of shell. In Fig.2 (b), rutile phase with a tetragonal crystal system and a space group $P4_2/mnm$ was confirmed in the surface part of $TiO_2$ shell.

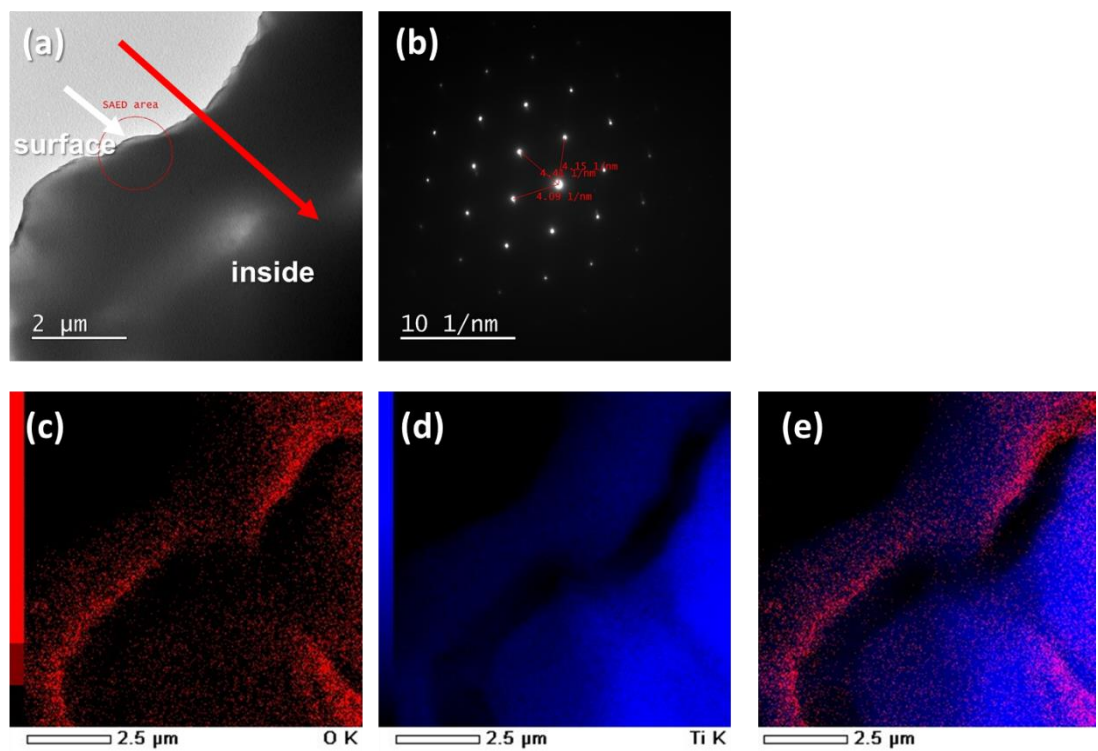

**Fig. 2.** (a) TEM image of obtained sample, (b) selected area electron diffraction pattern and EDS mapping the elemental distribution of (c) Ti, (d) O and (e) overlap

The chemical bonding state in synthesized $TiO_2$ was investigated by XPS and results are shown in Fig.S3. Fig.S3(a) displays XPS spectra of $Ti_{2p}$ orbital and peaks at 464.5, 462.8, and 461.1 eV can be assigned to $Ti^{4+}$, $Ti^{3+}$, and $Ti^{2+}$, respectively [17]. The peak area ratio of each Ti coordination number ($Ti^{4+}$, $Ti^{3+}$, and $Ti^{2+}$) were calculated by curve fitting using Gaussian function (Fig.S3 (b)). As a result, $Ti^{3+}$ was numerously existed on sample surface of synthesized $TiO_2$ which implies the formation of oxygen-deficient $TiO_2$ with efficient self-doping of $Ti^{3+}$ during MW synthesis. Furthermore, the spectra of O1s orbital showed obvious difference between the synthesized $TiO_2$ and the conventional one. The synthesized TiO2 had a shoulder peak at about 532 eV as shown in Fig. S4. The peaks at 530.1, 531.7 and 533.2 eV are assigned with $O^{2-}$ (Ti-O) and $OH^-$ and absorbed water [18,19]. In general, oxygen vacancy induce dissociation of $H_2O$, then taking place OH group chemisorption on the surface of $TiO_2$ [20,21]. Therefore, this result indicates that the synthesized $TiO_2$ has high concentration of oxygen vacancy.

   The UV-vis spectra and calculated Tauc plot of the synthesized $TiO_2$ are given by Fig.3. As a result, MW synthesized $Ti^{3+}$ doped $TiO_2$ exhibited significantly enhanced light absorption in the visible region with wavelength above 400 nm to compare with commercial $TiO_2$. The optical band gap of synthesized $TiO_2$ estimated from Tauc plot is 2.14 eV, which is much narrower than that of stoichiometric $TiO_2$. Such narrow bandgap can be attributed to the localized states at 0.75–1.18 eV below the conduction band minimum induced by oxygen vacancy [22]. As inserted photograph shown, the MW synthesized $Ti^{3+}$ doped $TiO_2$ shows grey-black color while stoichiometric (P-25) $TiO_2$ gives white color. It has been reported that electrons trapped in oxygen vacancy of $TiO_2$ can absorb light of a specified wavelength and then color becomes grey or black [23-25]. Fig.4 shows the photocatalytic activity of MW synthesized $Ti^{3+}$ doped $TiO_2$ characterized by degradation of Rhodamine B (RhB) under visible light irridiation. The synthesized $T^{3+}$ doped $TiO_2$ performed superior photocatalytic activity to commercial P-25 under visible light.

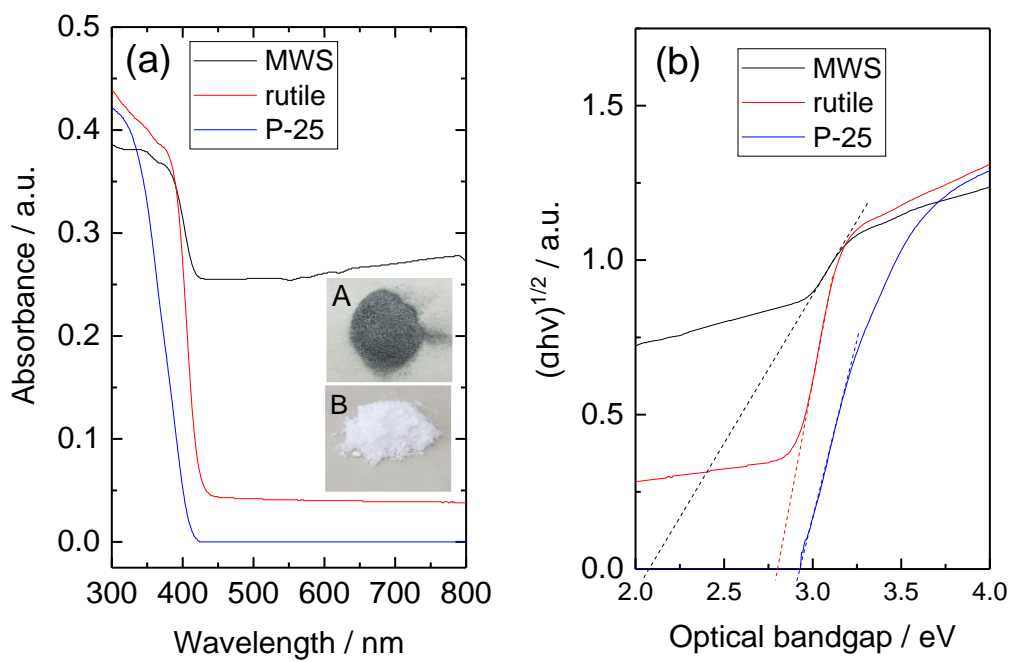

**Fig. 3.** Absorption spectra of raw and obtained material: (a) Absorbance and (b) Tauc plot. Inset pictures are MWS (A) and P-25 (B).

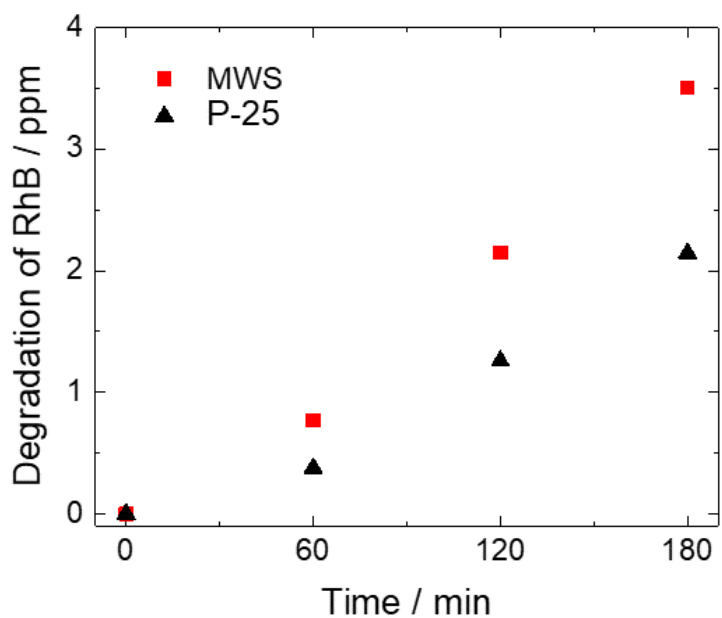

**Fig. 4.** Characterization of photocatalytic activity by degradation of RhB

## 4. Conclusions

In summary, $TiO_2$ was successfully synthesized via a one-step single-mode MW assisted process. Ti as target material selectively oxides in H-field throughout rapid heating, whose process requires less energy consumption and short time. The synthesized $TiO_2$ exhibits rutile phase according to the results of XRD and Raman spectrum. In addition, self-doping of $Ti^{3+}$ was confirmed in obtained $TiO_2$ by XPS measurement of $Ti_{2p}$ orbitals. Such MW synthesized $TiO_2$ showed excellent absorption property in visible light region with wavelength above 400 nm and narrow optical bandgap of 2.14 eV. In photocatalysis experiment of Rhodamine B degradation, photocatalytic performance of obtained $TiO_2$ was superior to commercially available $TiO_2$ under visible light irradiation. The innovative process for preparing metal oxide with specific chemical and physical properties open a brand-new strategy for developing functional materials.

**Acknowledgements:**

This work was supported by NITech Grant for Global Initiative Projects.